\documentclass[12pt]{article}
\usepackage{makeidx}
\usepackage{epsfig}
\usepackage{graphicx}
\usepackage{dcolumn}
\usepackage{bm}
\makeindex
\usepackage{amsfonts}
\usepackage{latexsym}

\def\b1{e^0}

\def\Tr{\mbox{Tr}}

\def\nn{\nonumber}
\def\aprime {\alpha^{\prime}}

\def\lesssim{\mathrel{\hbox{\rlap{\hbox{\lower4pt\hbox{$\sim$}}}\hbox{$<$}}}}
\def\gtrsim{\mathrel{\hbox{\rlap{\hbox{\lower4pt\hbox{$\sim$}}}\hbox{$>$}}}}

\def\u0{{\underline 0}}

\def\url{{\underline {r+\ell}}}

\newcommand{\be}{\begin{equation}}
\newcommand{\bea}{\begin{eqnarray}}
\newcommand{\ee}{\end{equation}}
\newcommand{\eea}{\end{eqnarray}}

\begin{document}
\begin{titlepage}
\bigskip
\rightline{} 
 \bigskip\bigskip\bigskip\bigskip
 \centerline{\Large \bf {Supergravity, M theory and Cosmology\footnote{
 To appear in the proceedings of Stephen Hawking's $60^{th}$
 birthday conference, Cambridge University, Jan. 2002}}}
 \bigskip\bigskip
 \bigskip\bigskip
 \centerline{\large Renata Kallosh}
 \bigskip\bigskip
 \centerline{\em Department of Physics, Stanford University, Stanford, CA 94305,
}
 \bigskip\bigskip
We discuss some recent attempts to reconcile cosmology with supergravity
and M/string theory.  First of all, we point out that in extended supergravities the  scalar masses are quantized in terms of the cosmological constant in de Sitter vacua:  the eigenvalues of the Casimir operator $3 m^2/\Lambda$  take integer values.
For the current value of the cosmological constant extended supergravities predict  ultra light scalars with the mass 
of the order of Hubble constant, $10^{-33}$ eV. This may have interesting consequences for cosmology. Turning our attention to cosmological implications of M/string theory, we present a possibility to use  string theory D-brane constructions to reproduce the main features of  hybrid inflation. 
We stress an important role  played by Fayet-Iliopoulos terms responsible for the positive contribution to the potentials and stabilization of moduli.

 \end{titlepage}

\section{Introduction}
Stephen Hawking has played an exceptional role in  life and scientific career of many people and I am one of them. My first encounter with Stephen was back in 1978 in Moscow. He told me that he knows  my work on  supergravity and  he would like to invite me to Cambridge for a supergravity workshop. Since at that time  I was never allowed to travel abroad, I did not believe that I will be able to do it.
However  Stephen Hawking's word had a strong weight for the Academy of Sciences of the USSR, and efforts of Academician M. Markov made my visit to  England possible. It is not clear to me how my career would evolve if not for this miracle.

Stephen Hawking's ability to be deeply involved in studies of gravity, cosmology as well as fundamental high-energy physics is extraordinary. We can see it clearly at this conference dedicated to his 60's birthday. We see people from these different scientific communities who mostly do not talk to each other  because the tools and lores are different. But all of us who came here  have some deep connection through various aspects of Stephen's work, as he made a strong impact on all of these fields. 

At present it is difficult to continue with traditional  developments in  supergravity and string/M theory. If any of these theories, to large extent based on supersymmetry, is  underlying a fundamental theory of gravity, one cannot ignore experimental facts established during the last few years. Recent cosmological observations based on the study of the anisotropy of the cosmic microwave background radiation (see  \cite{Sievers} for the most recent data) as well as supernova suggest that soon after the big
 bang our universe  experienced a stage of a very rapid accelerated expansion (inflation). Moreover, observations indicate that few billion years after 
the big bang  the universe entered a second stage of accelerated
 expansion. The rate of acceleration  now is many orders of 
magnitude smaller than during the stage of inflation in the 
early universe.

On the other hand, a significant progress in cosmology may require some input from supersymmetric M/String theory which at low energies are described by supergravities. It is plausible that supersymmetric particles will be discovered at high-energy accelerators. If this will happen, general relativity and cosmology will have to accommodate these facts.

In the meantime, as long as the results of the current cosmological experiments are not proven wrong and there is a hope that supersymmetry is present in nature, we may try to reconcile cosmology with supergravity and M/String theory.

\

De Sitter (dS) spaces relevant to  cosmological issues are extremely unnatural
for any supersymmetric theory. On the contrary, anti de Sitter spaces are natural spaces in fundamental theories with supersymmetry: in string theory, in M-theory and in most of supergravity theories.  Anti de Sitter spaces have unbroken supersymmetry, dS spaces always break supersymmetry.

It is difficult \cite{Gibbons:85} to construct dS vacua by compactification from ten or eleven dimensions, where the string theory and M-theory reign. 
The basic problem seems to originate from the  
compactification of M/string theory on internal space
 with the finite volume. Still some  
4d extended gauged supergravity theories  are known to have 
dS solutions with spontaneous breaking of supersymmetry. The first model of this kind was discovered in
\cite{Gates:1983ct} and many more were found later.
 These versions of 4d supergravity are related to 11d 
supergravity with a non-compact internal 7d space 
\cite{Hull:1988jw}.  

In this talk I will  discuss topics on which I  worked  or which are closely related.  I will not be able to cover all attempts to construct viable cosmological theories in supergravity and M-theory, see however, other talks at this conference. 

In the first part of the talk I will discuss   
extended supergravities $N\geq 2$ with dS vacua\footnote{Unextended N=1 supergravities have been studied in the cosmological context before. They are much less restrictive  and more difficult to relate to M/String theory.}, and their properties. My main statement is that in all known extended supergravities  {\it masses of scalars  in dS vacua are quantized in terms of the  cosmological constant}. 
 The  quantization condition is of the form 
\be
{m^2\over H^2}=k \,   
\label{quantization}\ee
Here $ \Lambda=3 H^2$ (in units with  $M_{P}=1$) is a cosmological constant equal to the value of the potential in dS extremum, 
 { \it $k$ are some  integers  of the order 1, completely independent of all parameters of the theory}. They may take negative values in models with tachyons.
I will point out here that   ${m^2\over H^2}$ is an eigenvalue of the Casimir operator  in dS space. Therefore the mass quantization condition described above has  geometric group-theoretic interpretation: {\it  in  extended supergravities with dS vacua eigenvalues of the Casimir operator take integer values}.

This part of the talk is based on work performed with Linde, Prokushkin and Shmakova  \cite{Kallosh:2001gr} where a large class of N=8,4,2 supergravities with dS vacua were studied.  Our purpose was to study the properties of the potentials near dS vacua which would be interesting either for early universe inflation or for the present day acceleration. We have found two important features of dS vacua: 

\begin{itemize}

\item in all theories of 4d extended supergravities which were known at that time, there are {\it tachyons in dS vacua} 

\item in all these theories {\it masses of scalars  in dS vacua are quantized in terms of the  cosmological constant}. We have found that  $k=3m^2/ \Lambda$ is an integer.

\end{itemize}

In \cite{Kallosh:2001gr} we had examples\footnote{It was pointed to us by A. Van Proeyen that one class of  N=2 supergravity models \cite{deWit:1984pk} was not analysed in \cite{Kallosh:2001gr}. We have checked now that these models also have integer values of Casimir operator,  $k= 12,-2, -6$. }  with $k=12, 6, 4, -6$, see eqs. (52-53) in  \cite{Kallosh:2001gr}. Quite recently new models of N=2 supergravity were constructed which have dS vacua without tachyons,  $m^2\geq 0$, \cite{Fre:2002pd}. A surprising feature of these theories is that again near dS extremum all  masses of scalar particles are quantized in terms of the  cosmological constant with  $k= 6, 4, 3,  2, 0$. 

In application to present cosmological constant this leads to immediate conclusion that there are ultra light scalars with the mass of the order
$$
m \sim H\sim 10^{-33}eV \ .
$$
The significance of this fact and the possibility to use these supergravity models in cosmology still  have to be understood.  The existence of such ultra light fields may be a desirable feature  for the description of the accelerated universe \cite{Carroll:1998zi}.  The presence of ultra light scalars signals that the corresponding potentials are very shallow. As we will see, in extended supergravities ultra light fields necessarily come in a package  with ultra small cosmological constant. Supersymmetry in dS vacua is broken spontaneously due to the presence of the cosmological constant, the scale of susy breaking here is $\sim 10^{-3} eV$. In this model, before it is coupled to a `visible sector',  both the small value of the cosmological constant as well as the ultra light masses of scalars are protected from large quantum corrections. The major problem is of course how to couple these theories  to the rest of the world. If they  play a role of a   `hidden sector', one may wonder whether  its properties,  the tiny cosmological constant and masses of the order of a  Hubble constant, will be preserved after coupling to the `visible sector'. We see now that preservation of the small cosmological constant may imply preservation of small scalar masses. Thus, extended supergravities suggest a new perspective for investigation of the {\it cosmological constant problem, intertwined with ultra-light scalars}.\footnote{These considerations were  stimulated by discussions with  Kaloper and  Linde.}

In the second part of the talk, based on the work done in collaboration with Dasgupta, Herdeiro and Hirano \cite{Herdeiro:2001zb}, \cite{Dasgupta:2002ew},
  we will pursue another strategy to reconcile superstring theory with cosmology. We suggest a particular possibility to use the D-branes of string theory in the cosmological context. The main feature of this development is a possibility to reinterpret various stages of hybrid inflation \cite{Linde:1991km}-\cite{Kallosh:2001tm} in  string theory context.  

We start with familiar well understood BPS objects, like D4 and D6 branes or 
D3  and D7 branes placed at some distance from each other.  Each of them separately is supersymmetric and stable.  Some {\it instability of the initial state of these branes is introduced 
via a  small deviation of supersymmetry for the system of two such objects}.  For example, for  D4/D6 model we use a {\it small angle between BPS branes} as a source of instability. Alternatively, in the dual model
one can  allow a {\it small magnetic field} to live on D7 brane so that D3 brane is attracted to it.
 This generates  expansion of the universe in a nearly dS 
 state, when the brane configuration is coupled to gravity. At some distance between the branes, the masses of the open strings become tachyonic and the process of symmetry breaking, resembling   the  tachyon condensation in string theory \cite{Sen:1998sm}, brings the system  towards the exit of inflation. The final stage in all cases we study is supersymmetric. In D4/D6 model the branes reconfigure and restore supersymmetry. The final stage is particularly nicely described in D3/D7 model: there is a bound state of D3/D7 system in which D3 dissolves on D7 as an instanton.  Embedding of the cosmological D-brane construction into 11d M-theory will be also discussed.

\section{Extended Supergravities with dS vacua}

The observation of Ref. \cite{Kallosh:2001gr} that there are tachyons in all supergravity models with $N\geq 2$ with dS vacua was made a while ago. 
I reported the results about tachyons and about the mass quantization  at the  conferences dedicated to 60'th birthday of John Schwarz and 25'th birthday of Supergravity. 
The  mass quantization observed in \cite{Kallosh:2001gr} (with  models \cite{deWit:1984pk} added)
\be
  {m^2\over H^2} =k   , \qquad k=12, 6, 4, -2, -6
\label{quantizationTach}\ee
remained a mystery.  After intensive discussions with practically all experts in the field we  concluded that
there is no reason to believe that the presence of tachyons in dS vacua is necessary in extended supergravities. Therefore one could expect that  new versions of supergravity without tachyons in dS space can be found. The art of constructing gauged supergravities is rather complicated. However new developments in cosmology urge us to find new gauged supergravities. Only when extended supergravities are gauged, they have non-trivial potentials.

Recently new studies of gauged supergravities were performed \cite{Gibbons:2001wy}
 and it was  also shown that  one may find new types of gaugings not studied before \cite{Andrianopoli:2002mf}. More importantly, new class of N=2 models with dS vacua without tachyons were constructed by    Fr\'{e},  Trigiante and  Van Proeyen, \cite{Fre:2002pd}.
A fascinating property of scalar masses quantized in terms of cosmological constant in dS vacuum remains true for all  models constructed so far. One finds that
\be
  {m^2\over H^2} =k   , \qquad k=6,4,3, 2, 0.
\label{quantizationT}\ee
Here we would like to describe  this quantization rule for models discussed in Refs. \cite{Kallosh:2001gr,Fre:2002pd} in geometric and group-theoretic terms.

The dS hypersurface embedded in 5d space with the flat metric is
\be
\eta_{\alpha \beta } X^\alpha X^\beta = - H^{-2},  \qquad \eta_{\alpha \beta }= {\rm diag}(1,-1,-1,-1,-1)\ .
\ee
The   Casimir operator $C_2$ is given by
\be
C_2 = -{1\over 2}  M^{\alpha \beta}M_{\alpha \beta} \qquad \alpha=0,i; \quad  i=1,2,3,4 \ , 
\ee
where the ten generators
\be
M_{\alpha \beta} =-i \left(X_\beta {\partial\over \partial X_\alpha }- X_\alpha {\partial\over \partial X_\beta } \right )
\ee
form the  $SO(1,4)$ algebra and $C_2$ commutes with all of them. Therefore $C_2$ is a constant in each representation. Our quantization means that it has integer eigenvalues in extended supergravities.
\be
\langle C_2 \rangle= \langle -{1\over 2}  M^{\alpha \beta}M_{\alpha \beta}\rangle = {m^2\over H^2}=k \ .
\ee
To explain this we will employ the dS studies used before by Dixmier \cite{Dixmier} to classify the representations in dS space. For $  k \geq {9\over 4}$ the representations are called principal series. For $0< k < {9\over 4}$ the representations are called complementary series. Finally $k=0$ belongs to discrete series of representations. The meaning of $k$ is nicely explained in \cite{ Gazeau:1999mi}. One starts with dS space with bounded global coordinates  describing a compactified version of dS space, a Lie sphere. 
The metric is 
\be
ds^2= {1\over H^2 \cos^2 \rho}\left (d\rho^2- d\alpha^2 -\sin^2\alpha d\theta^2 - \sin^2 \alpha \sin^2 \theta d\phi^2 
 \right)\ .
\label{dSmetric}\ee
Here $\rho$ is a timelike coordinate $-\pi/2< \rho <\pi/2$ and the rest of coordinates describes an $S^3$ part of the Lie sphere\footnote{The metric in (\ref{dSmetric})  is the standard one in the textbook \cite{Birrell:ix}, eq. (5.67) with $\rho= \eta-{\pi/2}$.}.

The massive scalar field equation in dS background is 
\be(\Delta_{LB} + m^2)\phi=0 \ ,
\label{scalar} \ee
 where the Laplace-Beltrami
operator on dS space in coordinates specified above is proportional to the Casimir operator $C_2$
\be
\Delta_{LB}= {1\over \sqrt {g}} \partial_\nu \sqrt {g}  g^{\mu\nu} \partial_\mu = H^2 \cos^2 \rho (\Delta_1- \Delta_3) = - H^2 C_2\ .
\ee
Here  $\Delta_1= \cos^2\rho
{\partial\over \partial\rho} \left( {1\over \cos^2\rho} {\partial\over \partial\rho}\right)$ and $\Delta_3$ is the Laplace operator on the hypersphere $S^3$. Thus the second order differential operator $C_2$ is
\be
C_2= \cos^2 \rho (\Delta_3- \Delta_1)= \sum_i (M_{0i})^2-  \sum_{i>j} (M_{ij})^2
\ee
The scalar field equation of motion (\ref{scalar}) can now be rewritten as\footnote{The standard equation for the scalar field in \cite{Birrell:ix}, eq. (5.68), is for a scalar field $\tilde \phi= \Omega^2 \phi$ after a conformal transformation to a frame with $d{\tilde s}^2= d\eta^2+\dots$. In this  frame the interpretation of $k$ as an eigenvalue of the Casimir operator is lost since the second order differential operator $C_2$ does not commute with the conformal transformation and therefore
$C_2 \tilde \phi \neq   k \tilde \phi $.}
\be
{1\over H^2}(\Delta_{LB} + m^2)\phi =0 \ , \qquad \Rightarrow \qquad C_2\phi = {m^2\over H^2} \phi
\ee 
This group-theoretical basis for quantization sheds a light on the universality of the quantization condition (\ref{quantization}). Still we do not have a derivation of the quantization condition (\ref{quantization}) in extended supergravity but only the observation that it takes place in all known at present models.

\

Here we will present a  truncated version of all   models in \cite{Fre:2002pd}, so that only axion-dilaton scalars are present. This will give us an opportunity to explain the mechanism for the appearance of the ultra light axions and dilatons in dS vacua with ultra small cosmological constant. The axion-dilaton system in string theory and ungauged supergravity  has $SL(2,R)$ or $SL(2, {\bf Z})$ symmetry. The action is given by
\be
{g}^{-1/2} L = {g^{\mu\nu}\partial_\mu \tau \partial_\nu \bar \tau \over (2 
\rm{Im }\,\tau)^2} \ .
\ee
A complex modular parameter $\tau= a-ie^{-\varphi}$ has some nice properties under  linear fractional transformations. These include in particular the shift of an axion $a\rightarrow a+\rm {const}$. The relevant part of new gauged N=2 supergravity action  is 
\be
 -{1\over 2}R + {1\over 4}[(\partial \varphi)^2 + e^{2\varphi}(\partial a)^2]-  V (a, \varphi)\ ,
\ee
where the potential is:
\be
 V= \Lambda \, [\cosh (\varphi-\varphi_0)+ {1\over 2} e^{(\varphi+\varphi_0)}(a-a_0)^2]\ .
\ee
We have presented the  axion-dilaton potential as a function of the cosmological constant $\Lambda$ and the critical values of the axion-dilaton field $\tau_0= a_0-ie^{-\varphi_0}$. At the minimum where $V'=0$, 
\be
\varphi=\varphi_0 \ , \qquad a=a_0 \ , \qquad V(a_0, \varphi_0)=\Lambda >0 \ .
\ee
The solution of equations of motion is a dS space with axion-dilaton field fixed at  constant values $\tau_0$. It is clear from the action that near dS vacuum the mass of the dilaton as well as the mass of the axion fields are equal to
\be
m_{\varphi}^2= m_a^2= 6 \,{\Lambda\over 3}=2\Lambda \ .
\ee
Note that the global axion shift symmetry $a\rightarrow a+\rm {const}$ is broken by the potential with the fixed value of $a_0$. 
 
The original parameters used in \cite{Fre:2002pd} are some particular combinations of gauge coupling $e_0$, Fayet-Iliopoulos term $e_1$  and magnetic rotation angles $\theta$, such that 
\be
 \Lambda = e_0 e_1 \sin \theta \ ,\qquad a_0= \cot \theta \ , \qquad e^{\varphi_0}= {e_0 \over  e_1 \sin \theta} \ .
\ee
In presence of other scalars there are few additional values $k=4,3,2$ 
as well as some flat directions with $m^2=0$.
This means that all of these scalars have a mass of the order $10^{-33}$ eV.

To the best of our understanding, no other theory so far had such a prediction. Interestingly, these ultra light masses occur in all newly discovered extended gauged supergravities  with dS minimum.

It is also interesting that this is the only known to us model of supergravity where  dilaton-axion stabilization naturally takes place, instead of the more familiar run-away behavior.

This result could have interesting cosmological implications. Indeed, in the early universe the light scalar fields may stay away from the minima of their potentials; they begin moving only when the Hubble constant determined by cold dark matter decreases and becomes comparable to $|m|$. A preliminary investigation shows that for $0<m^2<6H^2$ the fields move down very slowly and do not reach the minimum of the effective potential at the present stage of the evolution of the universe. This may result in noticeable changes of the effective cosmological constant during the last 10 billion years. Existence of this effect can be verified by  observational studies of acceleration of the universe. In addition, a slow drift of the fields to the minima of the effective potential may lead to a time-dependence of effective coupling constants.


\section{Hybrid Inflation with D-branes}

In the previous section I described properties of de Sitter state that can emerge in extended supergravity. Now we will try to see what may happen in M-theory describing interacting branes. But before we turn to M-theory, we will describe N=2 supersymmetric P-term hybrid inflation model \cite{Kallosh:2001tm}. Two different M-theory constructions that we suggested in \cite{Herdeiro:2001zb,Dasgupta:2002ew} are based on this model. 
 
\subsection{The potential of P-term hybrid inflation}

 Hybrid inflation scenario \cite{Linde:1991km}  can be
naturally implemented in the context of supersymmetric theories
\cite{Copeland:1994vg,Binetruy:1996xj}. The basic feature of
hybrid inflation is the existence of two phases in the
evolution of the universe: a slow-roll inflation in  dS
valley of the potential (the Coulomb phase of the gauge theory)
and a  tachyon condensation phase, or `waterfall stage', towards
the ground state Minkowski vacuum (a Higgs phase in gauge theory).

In $\mathcal{N}=1$ supersymmetric theories, hybrid inflation may
arise as F-term inflation \cite{Copeland:1994vg} or  D-term inflation
\cite{Binetruy:1996xj}. In $\mathcal{N}=2$ supersymmetric theories
there is a triplet of Fayet-Illiopoulos (FI) terms, $\xi^r$, where
$r=1,2,3$. Choosing the orientation of the triplet of FI terms,
$\xi^r$, in directions 1,2, F-term inflation was promoted to
$\mathcal{N}=2$ supersymmetry \cite{Watari:2000jh}. The more
general case with ${\mathcal N}=2$, when all 3 components of the
FI terms are present,  called \textit{P-term inflation}, was suggested in
\cite{Kallosh:2001tm}. When  $\xi^3$ is non-vanishing, a special
case of D-term inflation with Yukawa coupling related to gauge
coupling is recovered. In this fashion, the two supersymmetric
formulations of hybrid inflation are unified in the framework of
${\mathcal N}=2$ P-term inflation. This gauge theory has the
potential \cite{Kallosh:2001tm}
 \be V= {g^2\over 2}\Phi^\dagger\cdot
\Phi \, |\Phi_3|^2 - \left[{1\over 2} (P^r)^2 +P^r\left({g\over 2}
\Phi^\dagger \sigma^r \Phi+\xi^r\right)\right] \ ,\ee 
where $P^r$
is a triplet of auxiliary fields of the $\mathcal{N}=2$ vector
multiplet, $ \Phi$ is a doublet of  2 complex scalars, $\Phi_1$ and $\Phi_1$, forming a
charged hypermultiplet, $\Phi_3$ is a complex scalar from the
$\mathcal{N}=2$ vector multiplet and $g$ is the gauge coupling.
The auxiliary field satisfies the equation 
$
P^r= - (g\Phi^\dagger
\sigma^r \Phi/2+\xi^r)
$
and the potential for the choice $\xi^3=\xi$ simplifies to 
 \cite{Kallosh:2001tm} 
 \bea V=  {g^2\over 2} \left[(|\Phi_1|^2 +
|\Phi_2|^2) |\Phi_3|^2 + |\Phi_1|^2 |\Phi_2|^2+ {1\over
4}\left(|\Phi_1|^2 -|\Phi_2|^2+ \frac{2{\xi}}{g} \right)^2\right]\
, \label{pot} \eea
 which is depicted in Figure 1.
\begin{figure}[h!]
\centering \epsfysize=9cm
\includegraphics[scale=0.5]{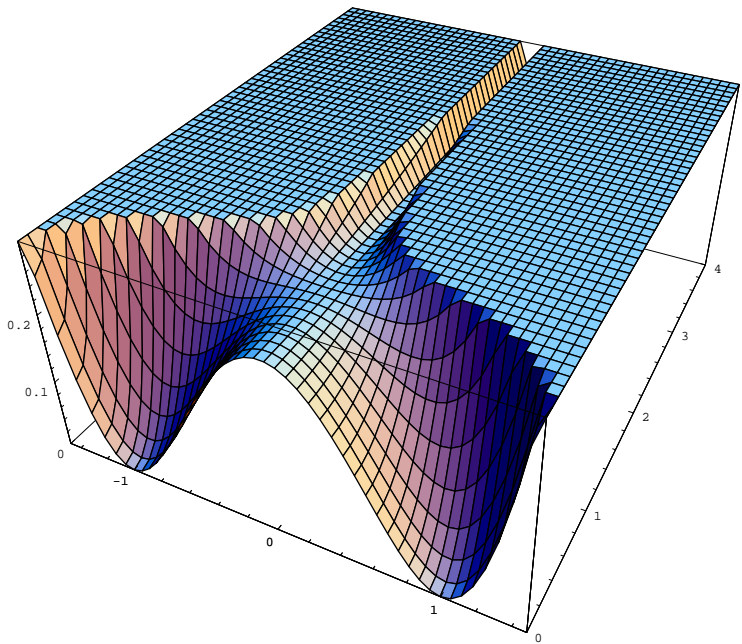}
\caption{Cosmological potential with Fayet-Iliopoulos term; notice
the dS valley is classically flat; it is lifted by a
one-loop correction, corresponding to the one-loop potential
between branes. In this figure the valley is along the
$|\Phi_3|$ axis; the orthogonal direction is a line passing
through the origin of the complex  $\Phi_2$ plane and we have put
$|\Phi_1|=0$. Notice there is no $\bf{Z}_2$ symmetry of the
ground state, it is just a cross section of the full $U(1)$
symmetry corresponding to the phase of the complex $\Phi_2$
field.} \label{}
\end{figure}
The potential (\ref{pot}) has a local
minimum, corresponding to a dS  space when coupled to
gravity, with $|\Phi_3|$ being a flat direction.
\begin{figure}[h!]
\centering \epsfysize=8cm
\includegraphics[scale=0.45]{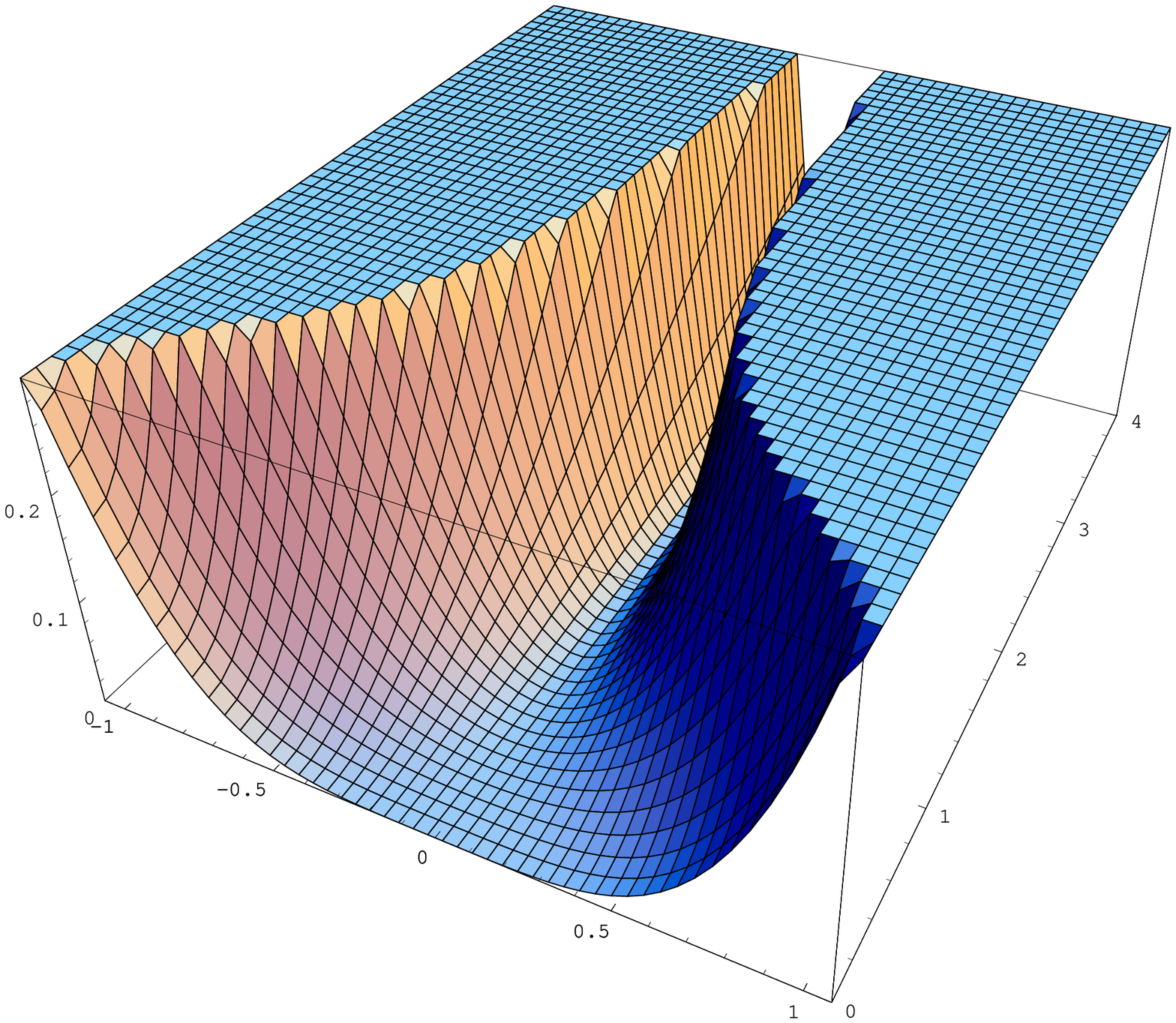}
\caption{Cosmological potential without Fayet-Iliopoulos term.} \label{pot1}
\end{figure}
This 
vacuum breaks all  supersymmetries spontaneously; here, the vev of
the hypers vanishes, $\langle \Phi_1 \rangle = \langle \Phi_2
\rangle =0$,  and the vev of the scalar from the vector multiplet,
which is the inflaton field, is non-vanishing, $\langle \Phi_3
\rangle \neq 0$. The masses of hypermultiplets  in the dS
valley are   split: \bea
M^2_{2}=g^2|\Phi_3|^2 - g{\xi}\ , \qquad M_{\psi} = g |\Phi_3| \ ,
\qquad M_1^2=g^2|\Phi_3|^2 +g {\xi}\ . \label{split}\eea Here
$\psi$ is the hyperino, $\Phi_1$ ($\Phi_2$) are positively
(negatively) charged scalars of the hypermultiplet. The value of
the potential at this vacuum is $V= {\xi}^2/ 2$. This is the
cosmological constant driving exponential expansion of the
universe.  The presence of the FI term breaks
supersymmetry spontaneously, which is imprinted in the fact that
the supertrace of the mass spectrum vanishes,
$
\mbox{STr}\, M^2\equiv \sum_{j}(-1)^{2j}(1+2j)M_j^2 =0$. The point where one of the
scalars in the hypermultiplet becomes massless, \bea M^2_{2}=
g^2|\Phi_3|_c^2 - g {\xi}=0 \ \  \Leftrightarrow \ \  |\Phi_3|_c =
\sqrt {\xi\over g} \eea is a bifurcation point. At $|\Phi_3|^2\leq
{\xi}/g$, the dS minimum becomes a dS  maximum;
beyond it, such scalars become tachyonic. The system is unstable
and the waterfall stage of the potential leads it to a ground
state.  Finally, the system gets to the absolute
minimum with vanishing vev for the scalars in the  vector
multiplet, $\langle \Phi_3 \rangle =0$, and non-vanishing vev for
the scalars in hypermultiplet,  $\langle \Phi_2 \rangle ^2=
2{\xi}/g$: Supersymmetry is unbroken. 
The
gauge theory one-loop potential lifts the flat direction, via a
logarithmic correction
\bea V = {1\over 2} {\xi}^2 + {g^2\over 16 \pi^2}
{\xi}^2 \ln { |\Phi_3|^2\over  |\Phi_3|_c^2}.
\label{1loop}\eea 
This is precisely what is necessary to provide a
slow roll-down for the inflaton field, since this is an attractive
potential. 
We would like to stress that in the absence of FI term $\xi$ none
of the interesting things takes place. The potential with $\xi=0$
is plotted in Figure 2. There is a Minkowski valley with the flat
direction.

\subsection{D4/D6 model}

In our first attempt\footnote{Earlier studies of brane inflation
were performed in  \cite{dvaliedi,BAB}.} to link string theory to
a gauge model with a hybrid potential \cite{Herdeiro:2001zb}, we
used a system with a D4-brane attached to heavy NS5-branes and having a
small angle $\phi$ relative
 to a supersymmetric position with regard to D6-brane. The initial state is slightly
non-supersymmetric and forces  D4 to move towards the D6-brane
(see also \cite{Kyae:2001mk}). This setup reproduces accurately
the properties of the non-supersymmetric dS  vacuum  of
P-term inflation. In particular, the mass splitting
of the scalars in the hypermultiplet is reproduced by the low-lying string states.
 Notice the inflaton $\Phi_3$ is the
distance in $4,5$ direction between D4 and D6 in the brane model.
\begin{figure}[h!]
\centering 
\includegraphics[scale=0.5]{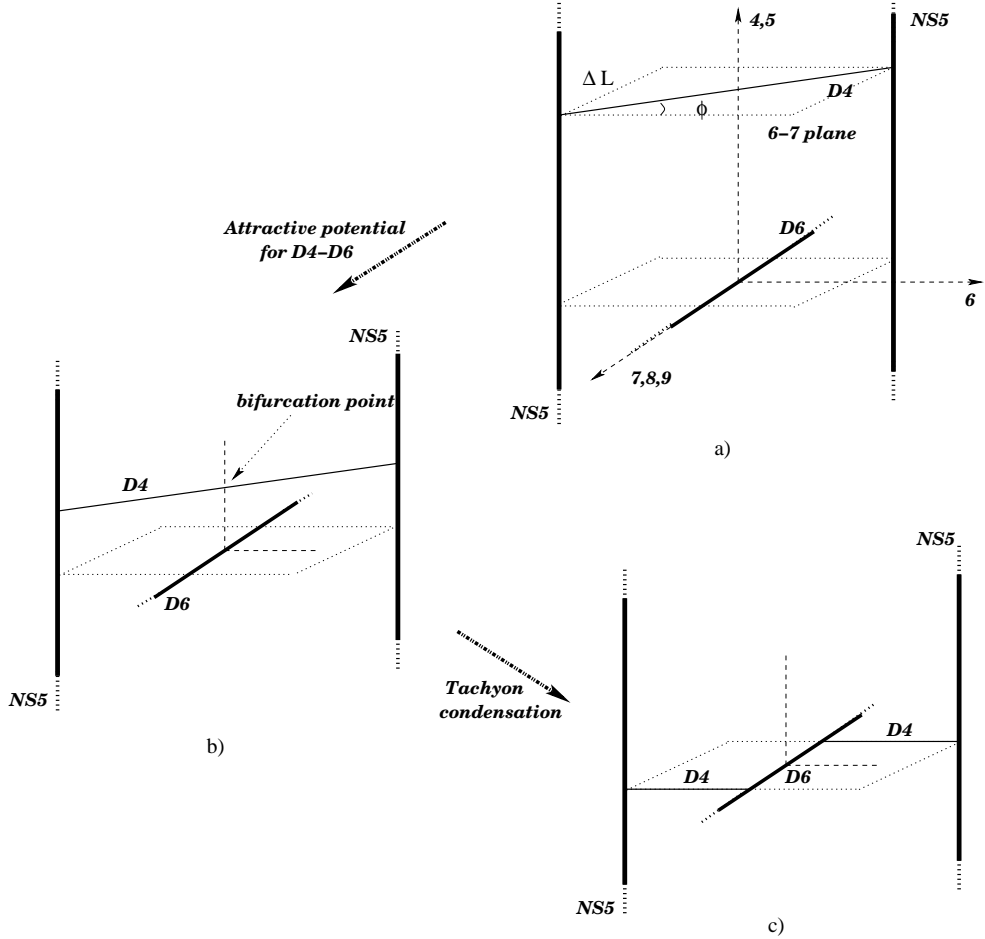}
\caption{Brane configuration evolution. a) For $\phi\neq 0$, supersymmetry is broken and D4-D6 experience an attractive force. b) At the bifurcation point, a complex scalar in the hypermultiplet becomes massless; when we overshoot  tachyon instability forms, taking the system to a zero energy ground state  shown in c) .}
\end{figure}
The most important dynamical effect is  the attractive potential between $D4$ and $D6$. In open string theory one-loop potential $V$ is given by
\begin{eqnarray}
 -\int_0^{\infty} {dt \over
t}\int_{-\infty}^{\infty} \frac{d^4p}{(2\pi)^4}\left[
\Tr_{NS}{1+(-1)^F \over 2}e^{-2\pi t L_{0}^{NS}}-\Tr_{R}{1+(-1)^F
\over 2}e^{-2\pi t L_{0}^{R}}\right]\ ,\nonumber
\label{bpot}
\end{eqnarray}
which results in
\begin{eqnarray}
V&=& \left({1 \over 8\pi^2\aprime}\right)^2 {\sin^2\phi \over
\cos\phi} \int_{\aprime/\Lambda^2}^{\infty} {dt \over t}
\exp\left[-2\pi t {(\Delta s)^2 \over \aprime} \right]
+{\cal O}(e^{-2\pi/t}), \nn\\
 &\sim& {g_{{}_{YM}}^2 \over 16 \pi^2 }\left(\frac{\tilde{\phi}}{2\pi \aprime}\right)^2
\log{(\Delta s)^2  \over \Lambda^2}.
\label{stringpot}\end{eqnarray}
Here $(\Delta s)^2$ is proportional $(x^4)^2+(x^5)^2$, the distance between branes. The string theory  potential  reproduces the one-loop correction in the field
theory (\ref{1loop}), including the numerical coefficient, in the
small angle and large separation approximation.
Therefore the motion of the $D4$ towards the $D6$ is the slow roll down of the inflaton. 

When the distance between the branes becomes smaller than the
critical distance  the spectrum of 4-6 strings develops a tachyon. The tachyon
condensation is associated to a phase transition. A final Higgs
phase with unbroken $\mathcal{N}=2$ supersymmetry is described in
this model by a reconfiguration of branes: D6 cuts D4 into two
disconnected parts, so that $\mathcal{N}=2$ supersymmetry is restored.

One attractive feature of D4/D6/NS5 model is that the deviation
from supersymmetry in the Coulomb stage can be very small and
supersymmetry is spontaneously broken. Nevertheless, a large
number of e-foldings can be produced within a D-brane.

The main difference between our model \cite{Herdeiro:2001zb}
and  other models of brane inflation
\cite{dvaliedi}-\cite{Burgess:2001vr}
is that our model provides the  brane description of the full potential
of hybrid P-term inflation. This includes both the logarithmic
quantum corrections to the Coulomb branch potential and the exit
from inflation with the corresponding supersymmetric Minkowski
vacuum.

\subsection{D3/D7 model}

The model suggested in \cite{Dasgupta:2002ew} describes a D3-brane parallel to a D7-brane at some distance. The distance again plays the role of the inflaton field. The supersymmetry breaking parameter is related to the presence of the antisymmetric
$\mathcal{F}_{mn}$ field on the worldvolume of the D7-brane, but
transverse to the D3-brane. When this field is not self-dual in
this four dimensional space, the supersymmetry of the combined
system is broken. This is (to some extent) a type IIB dual version
of the cosmological model proposed in \cite{Herdeiro:2001zb},
which guarantees that the good properties in the Coulomb stage are
preserved; in particular the spectrum and the attractive potential
should match the ones of P-term inflation. One immediate
simplification is that the NS5-branes are not needed any longer.
Also this model allows to address the issue of compactification and M-theory embedding.

The Higgs stage with 
\be P^r= - (g\Phi^\dagger
\sigma^r \Phi/2+\xi^r)=0
\label{ADHM}\ee
can be associated with the
Atiyah-Drinfeld-Hitchin-Manin construction of instantons with
gauge group $U(N)$. The moduli space of one instanton is the
moduli space of vacua of a $U(1)$ gauge theory with $N$
hypermultiplets and (\ref{ADHM}) is the corresponding ADHM
equation. We find an abelian non-linear instanton solution on the
worldvolume of the brane in the Higgs phase with associated
ADHM equation (\ref{ADHM}). Moreover, the presence of a
cosmological constant will translate as the resolution of the
small size instanton singularity.

One other nice feature of the D3/D7 cosmological model is that it
is well explained in terms of $\kappa$-symmetry of the D7-brane
action, both in Coulomb phase as well as in the Higgs phase. We  use in both cases the
BPS condition on the worldvolume  given by \cite{Bergshoeff:1997kr}
$$
(1-\Gamma)\,\epsilon =0. \label{susy} $$ 
Here $\Gamma(X^\mu
(\sigma),  \theta(\sigma), A_i(\sigma))$ is a generator of
$\kappa$-symmetry and it should be introduced into the equation
for unbroken supersymmetry with vanishing value of the fermionic
worldvolume field $\theta(\sigma)$. The existence or absence of
solutions to these equations  fits naturally in the two stages
of our cosmological model.

A perturbative  analysis of the string spectra is performed for a
 type IIB system with $D3$ and $D7$-branes plus a
constant worldvolume gauge field ${\cal F}$. 
This is illustrated in figure 4. 
\begin{figure}[h!]
\begin{picture}(75,0)(0,0)
\put(100,135){${\mathcal F}$}  \put(105,60){$D7$}
\put(130,0){$x^{4,5}$} \put(220,60){$D3$}
 \put(45,30){$x^{0,1,2,3}$} \put(115,25){$x^{6,7,8,9}$}
 \put(100,90){$_{\sigma=\pi}$} \put(245,112){$_{\sigma=0}$}
\end{picture}
\centering \epsfysize=13cm
\includegraphics[scale=0.5]{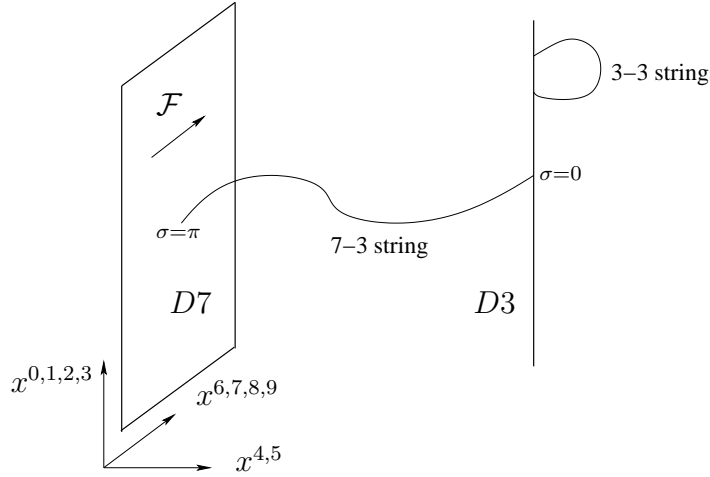}
\caption{The D3/D7 ``cosmological" system. The 3-3 strings give
rise to the ${\mathcal N}=2$ vector multiplet, the 7-3 strings to
the hypermultiplet and the worldvolume gauge field ${\mathcal F}$
to the FI terms of the $D=4$ gauge theory.}
\end{figure}
 There is a constant worldvolume gauge field
${\cal F}=dA-B$ present on D7:
\begin{equation}
 {\cal F}_{67}=\tan{\theta_1},~~~~{\cal F}_{89} = \tan{\theta_2} \
 , \label{fwv}
\end{equation}
responsible for spontaneous breaking of supersymmetry.  Note that if ${\cal F}$ is self-dual,
supersymmetry would be unbroken. This can
be explained via $\kappa$-symmetry. The lowest lying multiplet of states of open strings consist of
bosons whose masses are given by
\begin{equation}
M^2_{\pm}= {d^2\over (\pi \alpha')^2}\pm {\theta_1-\theta_2\over
2\pi \alpha'} \ ,
\end{equation}
where we observe that the boson of mass $M^2_{-}$ becomes
tachyonic at the critical distance.

The ${\cal F}$ field plays the role of the Fayet-Illiopoulos term,
from the viewpoint of the field theory living on the $D3$-brane.
It creates an instability in the system driving the $D3$-brane
into the $D7$-brane; this is  de Sitter stage. When the critical distance is reached a tachyon will form and the system will end in a
supersymmetric configuration.

D3/D7 bound state and unbroken $\kappa$-symmetry is specified by a solution of the equation $\Gamma\epsilon=\epsilon $.  
It  has an interpretation of
a D3-brane  dissolved into a D7-brane. The BPS condition requires that
 \be {{\cal F}_{ik}^-\over 1
+{\rm Pf}\, {\cal F}}= -  {B_{ik}^-\over 1 +{\rm Pf}\, B}\ .
\label{BPSframe}\ee 
Here ${\cal F}^-$ is the anti self-dual 2 form field on the world-volume of the D7 brane and $B$ is the anti self-dual 2 form of the background. This is the nonlinear instanton equation with known properties.

In the presence of the noncommutative parameter $\theta^-$ or
equivalently the $B^-$ field, the singularity of $U(1)$ instanton
is improved. In particular
the solution  has a finite instanton number.
Note that 
for $\theta^- = 0$  the instanton number would
have been divergent which corresponds to the fact that there is no Abelian
instanton with finite energy in absence of the non-commutativity parameter.

Thus the endpoint vacuum is described by a non-marginal bound state of D3 and D7-branes. The D3-branes on the D7-branes can be thought of as instantons on the D7-branes due to the Chern-Simons coupling,
$ {1\over 16 \pi^2}
\int_{D7}C_{4}\wedge F \wedge F= N \int_{D3}C_{4}.
$
 The Higgs phase of the D3/D7 system is actually identical to the noncommutative generalization of the ADHM construction of instantons. One may  argue therefore that our Minkowski vacuum is described by the noncommutative instanton \cite{Nekrasov:1998ss,Seiberg:1999vs}.
Thus we have found a possible link of the cosmological constant in spacetime ($0123$-space) and the non-commutativity in internal space ($6789$-space).


\section{M-theory on a four-fold with G-fluxes}

In addressing compactification of a D brane cosmological model to four
dimensions,  to recover four dimensional gravity, one is
 faced with the issue of anomalies. In many cases this
is associated to the requirement that the overall charge in a
compact space must vanish. Our
D3/D7 model can be considered within a more general setup, related to F-theory \cite{Vafa:1996xn}
 where the D7-brane charge is
cancelled by orientifold 7-planes or $(p,q)$ 7-branes. This seems
to provide a setup where, in string theory, the compactification
could be consistently performed\footnote{Some recent papers which
also consider orientifold planes to describe the inflationary
scenario are \cite{Burgess:2001vr}.}.

The above setup can be further simplified by lifting it to
M-theory.   Supersymmetry is broken by choosing a
non-primitive G-flux \cite{bbsis2}.  We may  summarise the situation in the following
table:
\begin{center}
 \begin{tabular}{*{9}{|c}|}
 \hline
 $\mbox {\bf Type~IIB}$ & ${\bf M-theory}$   \\
\hline $T^2/ (\Omega \cdot (-1)^{F_L} \cdot {\mathbb Z}_2) \times
K3$ & $T^4/{\mathbb Z}_2
\times K3$ \\
\hline
4($O7 + 4~D7$) & 4~orbifold~fixed~points \\
\hline
$D3$ & $M2$ \\
 \hline
${\cal F}$~on~$D7$ & Localised~G-flux~at~fixed-points \\
\hline
Coulomb phase & Non-primitive~G-flux \\
\hline
Higgs phase & Primitive~G-flux \\
\hline Away~from~orientifold~limit & $T^4/{\mathbb Z}_2 \ \ \rightarrow $ \ \ Smooth~$K3$ \\
\hline
 \end{tabular}
\end{center}
\begin{center}
 \end{center} The model that we are going to use is
M-theory compactified on a four-fold with G-fluxes switched on. To
get a ${\cal N} = 2$ theory we have to compactify M-theory on
$K3\times K3$. We shall take one of the $K3$ to be a torus
fibration over a ${\mathbb C}P^1$ base (see figure 5). 
\begin{figure}[h!]
\begin{picture}(75,0)(0,0)
\put(160,140){$T^2$} \put(170,90){${\mathbb C}P^1$} \put(235,
115){K3} \put (60,140){$(x^3,x^{10})$} \put(50,90){$(x^4,x^5)$}
\put(0,25){$(x^6,x^7,x^8,x^9)$} \put(190,25){K3}
\end{picture}
\centering \epsfysize=13cm
\includegraphics[scale=0.6]{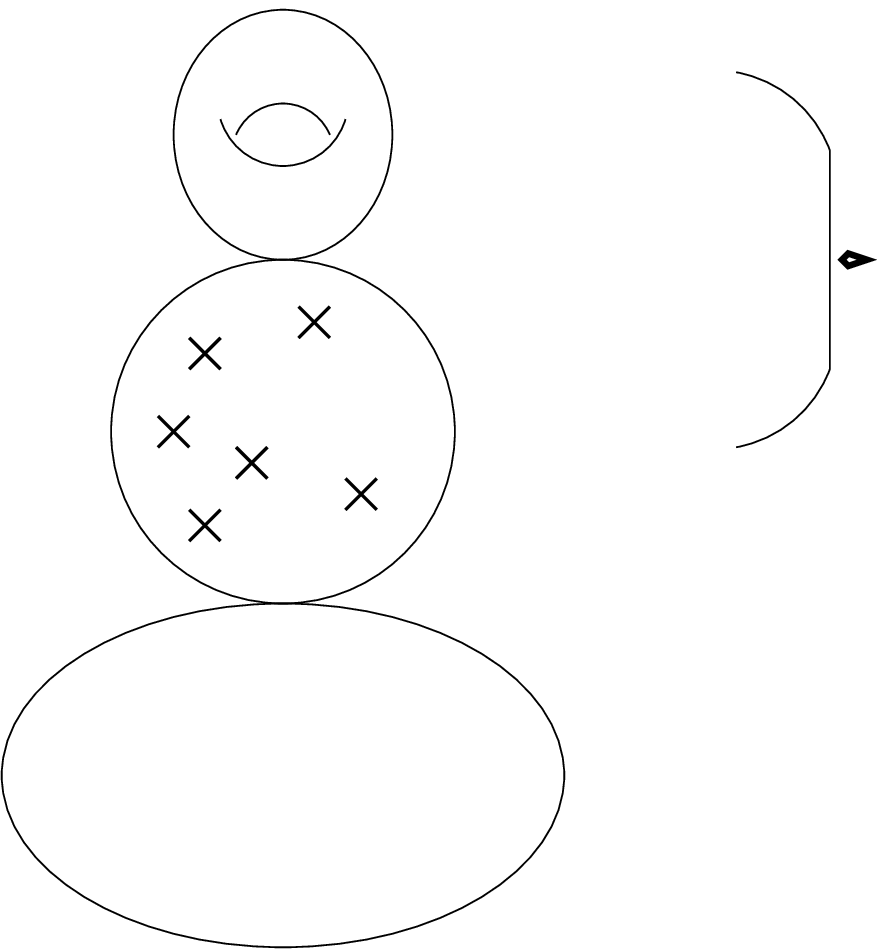}
\caption{The ``snowman" fibration of the $K3\times K3$ four-fold.
The crosses indicate points on the ${\mathbb C}P^1$ basis at which
the fibre tori degenerate. In the orbifold limit of the `top' K3
there will be four such points.}
\end{figure}
To go to type IIB theory we have to shrink the fiber torus to zero
size. 
We
assume the orbifold limit of K3 is the initial stage of our dynamical process. In
terms of type IIB language, we place a D3-brane at the center of
mass of the $4\times(4D7/O7)$ setup on one side of the ``pillow"
$T^2/{\mathbb Z}_2$ and another on its diametrically opposite
side, as in figure 6. We turn on the same gauge fluxes on all of
the four fixed points. The logarithmic potential 
creates a force between each pair of D3 and D7 branes, if we
assume the size of the $T^2/{\mathbb Z}_2$ to be large enough.
However in this configuration the total force between D3 and D7
branes is balanced, and the logarithmic potential being
approximately flat
 leads to a nearly de Sitter evolution. Quantum fluctuations will
destabilize the system allowing both the D3-brane to move towards
some D7-brane and the D7-branes to move away from the orbifold
fixed points. This is the beginning of the Coulomb phase or
equivalently the inflationary stage (figure 7).  The whole system eventually moves away from the orientifold limit,
which in M-theory language means we have a generic K3. Finally, we
expect the D3 to fall into one particular D7-brane as a
non-commutative instanton. This is the
supersymmetric configuration that has been studied in M-theory.
\begin{figure}[h!]
\begin{picture}(75,0)(0,0)
\end{picture}
\centering \epsfysize=13cm
\includegraphics[scale=0.5]{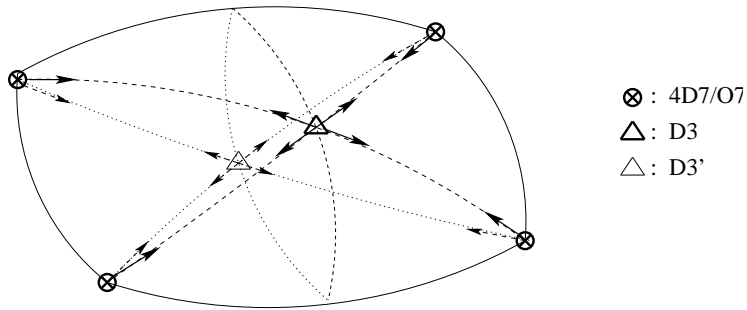}
\caption{The initial brane configuration on the ``pillow"
$T^2/{\mathbb Z}_2$.}
\end{figure}
\begin{center}
\begin{figure}[h!]
\begin{picture}(75,0)(0,0)
\end{picture}
\centering \epsfysize=13cm
\includegraphics[scale=0.5]{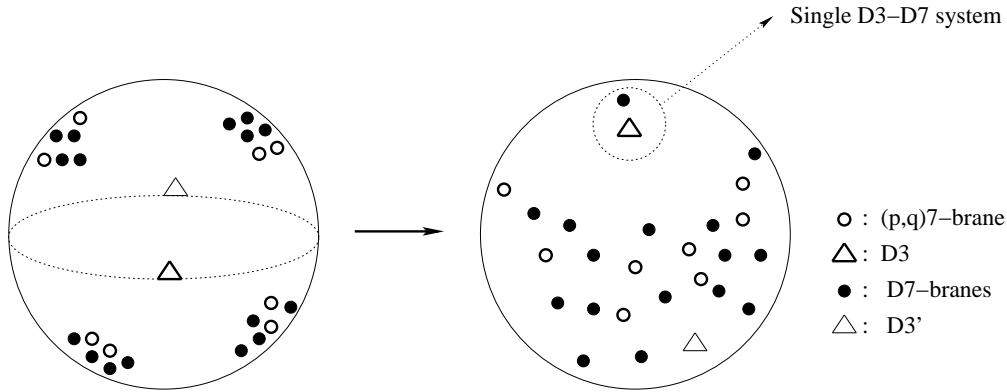}
\caption{The Coulomb branch. As the system is driven away from the
unstable point of figure 2, $T^2/{\mathbb Z}_2  \ \rightarrow \
{\mathbb C}P^1$, the orientifold planes split into $(p,q)$
7-branes, and the D3-brane will eventually fall into one D7-brane
as an instanton.}
\end{figure}
\end{center}
The 4-fold vacua has a tadpole anomaly given by $\chi /24$
where
$\chi$ is the Euler characteristics of the 4-fold.  If $\chi /24$ is
integral, then the anomaly can be cancelled by placing a sufficient
number
of spacetime filling M2-branes $n$
 on points of the compactification manifold. There
is also another way of canceling the anomaly and this is through
G-flux. The G-flux contributes a $C$ tadpole through the
Chern-Simons coupling $\int C\wedge G \wedge G$. When  $\chi /24$
is not integral then we need both the branes and the G-flux to
cancel the anomaly. The anomaly cancellation formula becomes
\begin{equation}
{\chi \over 24} = {1\over 8\pi^2}\int G \wedge G + n \ ,
\end{equation}
which must be satisfied for type IIA or M-theory. The anomaly cancellation condition will now become, 
$
{\chi \over 24}= n + \int H_{RR} \wedge H_{NSNS} \ ,
$
in type
IIB
theory, where $n$ is the number of D3 branes \cite{drs}.
Thus we have suggested here a
compactified picture in M-theory on $K3 \times K3$ manifold with a
choice of G-flux on it related to  a cosmological D brane models of P term inflation.

\

We have presented above an attempt to use in the context of cosmology the structures developed  over the years in M/string theory. Mostly they were used before in the BPS context with unbroken supersymmetry. Even the non-BPS branes were not used much for time-dependent cosmological evolution. Our analysis suggests that all of these tools that have been developed in M/string theory
can be very useful for finding realistic models of hybrid inflation.

\

In the list of open questions in the context of Supergravity, M theory and Cosmology the foremost important  are: Are these theories compatible? Can we learn from the fundamental  M/string  theory and supergravity about cosmology? There is no clear answer yet. In this talk I described some early attempts in which I have participated, to address these issues. One may expect much more to come in the near future.   Hopefully, the cosmological experiments as well as the high-energy experiments  will help to direct the developments of the theory. 

\

Happy birthday, Stephen!
\vskip .5cm
\centerline{\bf Acknowledgements}
It is a pleasure to thank J. Gates, G. Gibbons, B. de Wit, S. Ferrara, P. Fr\'{e}, D. Freedman, C. Hull, N. Kaloper,  L. Kofman, M. Ro\v{c}ek, M. Trigiante, A. Tseytlin, A. Van Proeyen and N. Warner for useful discussions of supergravity-cosmology issues,  and my collaborators on projects described above, K. Dasgupta, C. Herdeiro, S. Hirano, A. Linde, S. Prokushkin, M. Shmakova. 
Also
I want to thank the organizers of the Future of Theoretical Physics and
Cosmology conference, for a very stimulating meeting. This work
was supported by NSF grant PHY-9870115.
\printindex

\end{document}